\documentclass[journal]{IEEEtran}
\usepackage{graphics} 
\usepackage{amsmath} 
\usepackage{amssymb}  
\usepackage{comment}
\usepackage{authblk}
\usepackage{todonotes}
\usepackage{graphicx,caption}
\usepackage{mathtools}
\usepackage{comment}
\usepackage{algorithm,algpseudocode}
\usepackage{subfigure}
\usepackage{graphicx}
\usepackage{float}
\usepackage{amsfonts}
\usepackage{amsmath}
\usepackage{bm}
\usepackage{mathtools}
\usepackage{tabularx}
\usepackage{array}
\usepackage{breqn}
\usepackage{setspace}
\usepackage{supertabular}
\usepackage{multirow}
\usepackage{colortbl}
\usepackage{tabu}
\usepackage{hyperref}
\usepackage{longtable}
\newcolumntype{?}{!{\vrule width 2pt}}
\newcolumntype{H}{>{\setbox0=\hbox\bgroup}c<{\egroup}@{}}
\usepackage{url}
\newcolumntype{Y}{>{\RaggedRight\arraybackslash}X}

\DeclarePairedDelimiter\floor{\lfloor}{\rfloor} 
\usepackage[export]{adjustbox}
\begin{document}
\title{\LARGE \bf
An Asynchronous, Decentralized Solution Framework \\ for the Large Scale Unit Commitment Problem
}
\author{\IEEEauthorblockN{Paritosh Ramanan\IEEEauthorrefmark{1}\IEEEauthorrefmark{2},Murat Yildirim\IEEEauthorrefmark{3}, Edmond Chow\IEEEauthorrefmark{1} and Nagi Gebraeel\IEEEauthorrefmark{2}}
\thanks{\IEEEauthorrefmark{1}School of Computational Science and Engineering, Georgia Institute of technology, Atlanta, GA, USA 30332}
\thanks{\IEEEauthorrefmark{3}College of Engineering, Wayne State University, Detroit, MI, USA 48202}
\thanks{\IEEEauthorrefmark{2}H. Milton Stewart School of Industrial and Systems Engineering, Georgia Institute of technology, Atlanta, GA, USA 30332.\\paritoshpr@gatech.edu,murat@wayne.edu,echow@cc.gatech.edu,\\nagi@isye.gatech.edu}

}

\IEEEpubid{\begin{minipage}{\textwidth}\ \\ \\
\\[12pt]
DOI 10.1109/TPWRS.2019.2909664\\
\copyright 2019 IEEE.  Personal use of this material is permitted.  Permission from IEEE must be obtained for all other uses, in any current or future media, including reprinting/republishing this material for advertising or promotional purposes, creating new collective works, for resale or redistribution to servers or lists, or reuse of any copyrighted component of this work in other works.
\end{minipage}} 

\markboth{Accepted to IEEE Transactions on Power Systems}{}

\maketitle


\begin{abstract}
With increased reliance on cyber infrastructure, large scale power networks face new challenges owing to computational scalability. In this paper we focus on developing an asynchronous decentralized solution framework for the Unit Commitment(UC) problem for large scale power networks. We exploit the inherent asynchrony in a region based decomposition arising out of imbalance in regional subproblems to boost computational efficiency. A two phase algorithm is proposed that relies on the convex relaxation and privacy preserving valid inequalities in order to deliver algorithmic improvements. Our algorithm employs a novel \textit{interleaved binary} mechanism that locally switches from the convex subproblem to its binary counterpart based on consistent local convergent behavior. We develop a high performance computing (HPC) oriented software framework that uses Message Passing Interface (MPI) to drive our benchmark studies. Our simulations performed on the IEEE 3012 bus case are benchmarked against the centralized and a state of the art synchronous decentralized method. The results demonstrate that the asynchronous method improves computational efficiency by a significant amount and provides a competitive solution quality rivaling the benchmark methods.
\end{abstract}

\vspace{-10mm}

\section*{Nomenclature}
\vspace{-5mm}
\textbf{Sets}:
\vspace{-5mm}
\begin{center}
\begin{tabular}{c lc}
$\mathcal{R}$ & The set of all regions\\[1mm]
$\mathcal{N}_r,G_{r},\mathcal{U}_{r},\mathcal{V}_{r},\mathcal{I}_{r}$ & Neighboring regions, generators,\\& boundary,  foreign and internal buses of \\& region $r$\\[1mm]
$\mathcal{B}_r$ & $\mathcal{U}_r \cup \mathcal{V}_{r}$, Boundary, foreign buses of $r$\\[1mm]
$\mathcal{N}^b_r$ & Neighboring regions connected to  \\ & bus $b \in \mathcal{U}_{r}$\\[1mm]
$G^b_{r},\mathcal{U}^b_r,\mathcal{V}^b_{r},\mathcal{I}^b_r$ & Generators, boundary, foreign and \\& internal  buses connected to \\& bus $b \in \mathcal{U}_{r} \cup \mathcal{I}_r$\\[1mm]
$\mathcal{B}^b_r$ & $\mathcal{U}^b_r \cup \mathcal{V}^b_{r} \cup \mathcal{I}^b_r$, Neighboring buses of \\& bus $b$\\[1mm]
$T$ & Operational planning horizon
\end{tabular}
\end{center}

\textbf{Decision Variables} (at $t\in T$):
\vspace{-1mm}
\begin{center}
\begin{tabular}{c lc}
$y^{g}_t$ & The electricity dispatch of generator $g$\\[1mm]
$x^{g}_t \in \{0,1\}$ & The commitment decision variable of $g$\\[1mm]
\end{tabular}
\end{center}

\begin{center}
\begin{tabular}{c lc}
$\theta^{b}_{t}$ & The phase angle at bus $b$\\[1mm]
$\tilde{\theta}^{b,r' }_t$ & The phase angle of bus $b$
where $b \in \mathcal{U}_{r'}$ \\ & and $r' \in \mathcal{N}_r$\\[1mm]
$f^{uv}_{t}$ & Power flow from bus $u$ to $v$ such that $u \in \mathcal{U}_{r}$ \\& and $v \in \mathcal{V}^u_{r}$\\[1mm]
$\pi^g_{Ut}, \pi^g_{Dt} $ & The up and down variable of generator g\\[1mm]
$p_{r,t}$ & Production difference at region $r$\\[1mm]
$\lambda^{b}_t$ & The Lagrangian multiplier with respect to \\&phase angles of bus b where $b \in \mathcal{U}_{r} \bigcup \mathcal{V}_{r}$\\[1mm]
$\phi^{uv}_t$ & The Lagrangian multiplier with respect to flow\\& from bus $u$ to bus $v$ where $u \in \mathcal{U}_{r}$ and $v\in \mathcal{V}_{r}$\\& for any region $r$\\[1mm] 
$\eta_{t}$ & The Lagrangian multiplier with respect to \\
  &production difference of region $r \in \mathcal{R}$\\[1mm]
\end{tabular}
\end{center}

\textbf{Constants}:
\vspace{-2mm}
\begin{center}
\begin{tabular}{c lc}
$d^{g},c^{g},S^g_U,S^g_D$ & The dispatch cost, commitment cost\\& start-up cost shut-down cost of generator $g$\\[1mm]
$P^{g}_{min}, P^{g}_{max}$ & Minimum and maximum capacity of $g$\\[1mm]
$M^g_U, M^g_D, R^g$ & Minimum up time, down time and ramp-up \\& ramp-down constant for $g$\\[1mm]
$\delta^{b}_t$ & The demand at bus $b$ at $t \in T$\\[1mm]
$F^{uv}_{max}$ & Maximum capacity of line  connecting  buses \\& $u$ and $v$  such that $u \in \mathcal{U}_{r}$ and $v \in \mathcal{V}^u_{r}$\\[1mm]
$\rho_{\theta},\rho_{f},\rho_{p}$ & Penalty parameter for phase angles, flows and\\& production difference \\[1mm]
$\Gamma^{uv}$ & Phase angle conversion for line $uv$\\[1mm]
\end{tabular}
\end{center}


\section{Introduction}\label{sec:intro}
Unit Commitment (UC) in power networks is a well-studied optimization problem that determines the optimal power generation schedule for a fleet of networked generators. Any UC solution framework can be broadly divided into two parts, the data component and the UC problem component. While the problem component consists of the solution methodology and the optimization problem formulation, the data component consists of infrastructural information pertaining to the network topology, transmission lines, buses and generators. UC is a computationally challenging problem due to its scale, discrete nature and the vast amount of data that is essential to obtain a solution. Usually, UC is solved in a centralized manner at a control center where the problem component must be co-located with the data component. 
A distributed UC solution relies mainly on parallel two-stage techniques in order to obtain faster solution times. Such techniques exploit the 
scope for parallelism among the two stages to provide computational speedup \cite{ucasync,kim1,kim2,hpcuc}.  However they fail to remove the constraint of co-locating the data and the problem components. The UC problem is a stepping stone towards more complex planning problems. Emerging applications of UC involving data-driven operations planning will not be amenable to a centralized model of computation. This will be particularly important as UC problem faces new challenges in data acquisition and interpretations in the areas of integration of renewables, incorporation of maintenance, transmission line switching and prevention of cascading failures. In such problems, co-locating data and problem components may prove to be infeasible.

\begin{figure}[!ht]
\begin{minipage}{0.5\textwidth}
\centering
\includegraphics[width=\textwidth,keepaspectratio]{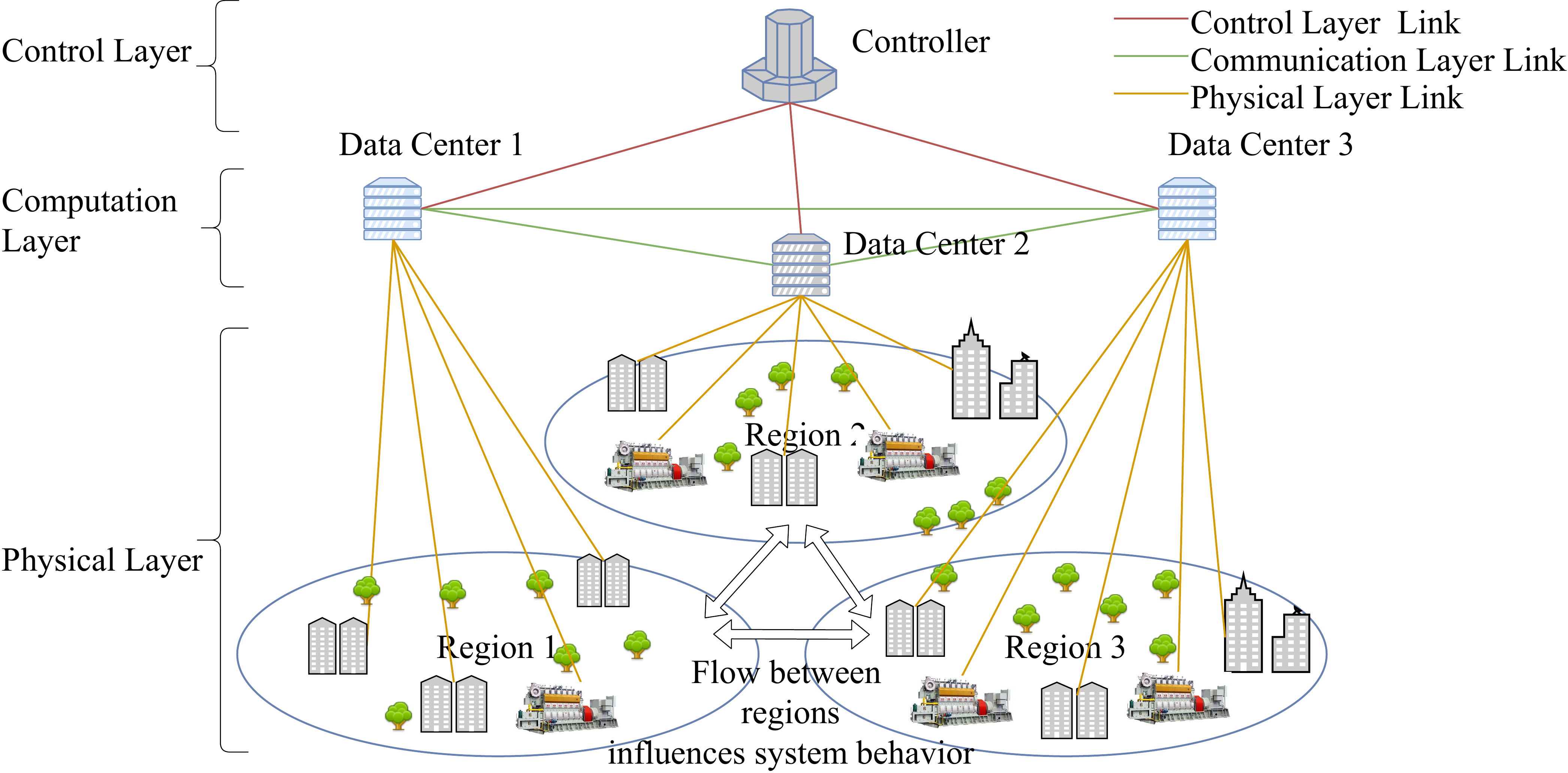}
\caption{A power network schematic for a decentralized framework}\label{fig:schematic}
\end{minipage}
\end{figure}

Unlike the distributed method, a decentralized solution framework is obtained by decomposing the global UC problem component into smaller local subproblems held by multiple computing agents. Each agent only holds a slice of the global data pertaining to its own local subproblem. Therefore, only the local subproblem and the corresponding network data needs to be co-located. The decentralized UC problem considered in this paper utilizes a region based decomposition strategy for a large scale power network. Every region is assigned to a unique computing agent denoted by a data center as shown in Figure \ref{fig:schematic}. The region based decentralized framework is laterally divided into three parts: physical, computation and control layers. The system level data comprising of network flow estimates is gleaned from the physical network layer by data centers located in the computation layer. These data centers, scattered across a wide geographic area are responsible only for their own local subproblems. At each region, local computation at the data centers yields new estimates of network flow variables that govern transmission lines. The updated estimates pertaining to shared transmission lines between two regions is communicated by the corresponding agents. The newly received network flow estimate is used as input to the next local computation step. The global progress of such an iterative scheme involving local computation and communication is tracked by the control layer. Eventually, the decentralized approach leads to balance of network flow estimates among the regions thereby signifying a solution to the global UC problem. 

A decentralized approach to the UC problem has numerous advantages. First, it improves computational efficiency since a region based decomposition of the global problem yields much smaller local subproblems that can be solved in parallel. Second, since each region operates independently, a decentralized method is highly suited for a geographically distributed computational architecture. Third, sensitive operational data can be held privately by each region, thus, limiting data sharing to specific parameters that do not violate privacy. Lastly, it has been found that a decentralized agent-to-agent communication is more efficient in terms of communication latencies \cite{assran_rabbat,ayetkin}.  A distributed agent-to-master  communication model introduces a single point of failure with the master bearing the heavy burden of processing information sent by all the workers. Since the master can only process information from one agent at a time, the other agents must wait till the master has had an opportunity to process and respond to their corresponding message resulting in poor scalability with respect to increasing problem sizes. Since communication is more expensive than computation \cite{comm_exp} idle time can be eliminated or significantly reduced by allowing agents to send messages to each other instead of waiting on one master to respond As a result decentralized communication improves computational performance.


However, one of the disadvantages of current decentralized methods is the need to fully \textit{synchronize} computation among all agents. In synchronous settings, all regions perform their local computational step, and wait for the other regions to finish before proceeding to the communication step. In a large scale decentralized power network, the expectation of a fully synchronous system can be highly misplaced \cite{inevitable}. Local subproblems might vary in their computational complexity owing to different problem sizes, which often leads to computational imbalance. The computing hardware employed by the various regions might be different, leading to an imbalance at the hardware level itself. In reality, a region's UC problem is localized to a data center within close geographical proximity to regional power assets. A practical solution to a large scale decentralized UC problem with a sizable number of regions would involve multiple data centers scattered over a vast geographical area thereby incurring significant communication costs. Therefore, in a real world implementation, a synchronous approach might suffer from significant idle time incurred due to local heterogeneity and increased communication costs leading to poor computational efficiency and slower progress.

An asynchronous approach has the potential to provide reliable, fast and robust decisions for power system optimization problems. In an asynchronous setting, all the regions perform their local updates based on the latest available information from their peers. Therefore, unlike the synchronous approach, the computational bottleneck arising from slower regions is eliminated, which minimizes idle time and improves computational efficiency. It naturally follows that an asynchronous method would also be resilient to the computational imbalance of local problems due to heterogeneous hardware and communication latencies. Further, an asynchronous decentralized method is regarded more favorably with respect to the mitigation of cyber-attacks since regional computations progress independently and the global objective remains unchanged. 

In this paper, we focus on developing a novel \textit{asynchronous decentralized computational framework} for solving large scale UC problems.  The main contributions of our paper can be summarized as follows:
\begin{itemize}
\item We develop a two phase asynchronous decentralized algorithm for solving the decentralized UC problem asynchronously. The algorithm iteratively solves the convex relaxation in the first phase. In the second phase, the binary constraints on the decision variables are imposed. We strengthen our two phase approach with privacy preserving valid inequalities that lead to sound solution quality and robust computational performance. 
\item We propose a novel \textit{interleaved binary} mechanism that allows regions to advance to the binary phase after having exhibited consistent local convex convergence behavior irrespective of global convex convergence. This also leads to a significant improvement in computational efficiency.  
\item We propose and implement the concept of a controller to facilitate two-way message exchanges at discrete global clock ticks among neighboring regions which satisfies a crucial convergence assumption \cite{woz}.
\item We develop a custom-made software framework based on the asynchronous reformulation that is fine tuned specifically for the UC problem. We present simulations with respect to the 75, 100 and 120 region scenarios of the IEEE 3012 bus system on a high performance computing environment using MPI semantics. 
\end{itemize}

A typical application of our method would be for a large scale ISO or vertically integrated power company. In such a scenario, participating regions could solve the global problem in a decentralized manner without revealing their infrastructural data while having superior computational performance with respect to centralized methods. Our algorithm is mainly for transmission level operators. However, our methodology is generic and could potentially be applied for coordinating transmission and distribution coordinated system operations as well.

The rest of the paper is divided as follows. Section \ref{sec:rw} we discuss various other approaches explored in literature with respect to decentralized, asynchronous methods.  In Section \ref{sec:DUC}, we present the region based decentralized decomposition of the UC problem. We present the asynchronous decentralized algorithm in Section \ref{sec:sm} based on \cite{woz} and develop a privacy preserving valid inequality that delivers algorithmic improvements. We discuss the implementation aspects of our algorithm and introduce the concept of a controller to successfully orchestrate two-way message exchanges as required by our algorithm. We present results from a robustness and benchmarking study in Section \ref{sec:results} and compare the asynchronous method against the centralized and synchronous variants. We conclude our paper in Section \ref{sec:conc}.
\vspace{-2mm}
\section{Related Works}\label{sec:rw}

Augmented Lagrangian techniques like the Alternating Direction Method of Multipliers (ADMM) \cite{admm_boyd} have been useful for solving decentralized constrained optimization problems with good convergence properties \cite{cons_wyin}. In the asynchronous optimization literature, the main focus has been on unconstrained optimization \cite{tsitsiklis1986distributed,liu2015asynchronous1,rabbat2014asynchronous}. Recently, there has been a growing literature on asynchronous constrained models to address a more general class of optimization problems. 
Chang \cite{chang2016proximal} proposes an asynchronous ADMM oriented solution for constrained optimization problems with time-varying networks and also under communication errors, whereas Eckstein \cite{eckstein2017simplified} proposes an asynchronous ADMM like method for multi-block decomposable problems suited for an HPC environment with shared memory capabilities and all-to-all connectivity among compute nodes. In contrast, Wei and Ozdaglar \cite{woz} propose an asynchronous ADMM algorithm for distributed constrained optimization that makes no assumptions about compute capabilities of the hardware or about the communication links present in the network. Further the authors show the convergence of their algorithm for distributed constrained optimization problems making their algorithm highly suited for a geographically distributed set of compute agents communicating over potentially high delay inducing links. 

Asynchronous master-slave type distributed optimization techniques have recently gained popularity within the power systems domain. 
Zhang and Kwok \cite{async-partial-ADMM} propose an ADMM implementation that takes advantage of partial progress being made by slaves with respect to the master, resulting in higher computational efficiency. Within the domain of power systems a similar idea is explored by Aravena and Papavasilou\cite{ucasync} in terms of a distributed asynchronous two stage stochastic UC model where the dual iterations and the feasibility recovery between the master and the slaves occurs asynchronously. Papavasilou \textit{et al.}\cite{hpcuc} propose a HPC solution framework for solving the stochastic UC problem with dual decomposition. Similar work has been done in the asynchronous domain by Kim \textit{et al.} \cite{kim1,kim2} exploiting the asynchrony arising out of load imbalance between the various slaves and the master in two-stage stochastic problems specifically for the security constrained UC and the stochastic UC problems. A limitation of these methods over the decentralized approaches, is the requirement of a master node. 
While offering much potential in terms of computational efficiency, the presence of a master problem with infrastructural data pertaining to the entire network can prove to be impractical in a real world setting owing to reasons mentioned in Section \ref{sec:intro}.

A hallmark of a decentralized solution is the absence of a master problem holding global data. In this domain, the work done by Feizollahi \textit{et al.} \cite{javad} provides a synchronous decentralized fix-and-release approach for large scale UC problems. However, in their framework, the UC fix-and-release pertaining to the MIP  may not be ideal for a geographically distributed computing environment. A direct asynchronous extension of the synchronous decentralized computational framework for UC presented in \cite{javad} has been explored in \cite{ramanan2017asynchronous,wang2018fully}. In a similar context, Guo \textit{et al.} \cite{guo2017asynchronous} provide an asynchronous decentralized method for non-convex problems in power systems, which has a similar computational outline as the previous two works but as pointed out in \cite{ramanan2017asynchronous}, such an approach might suffer from poor solution quality when it comes to MIP problems such as asynchronous decentralized UC. While these methods show great promise in computational efficiency, the solution quality arising from such types of asynchronous decentralized UC frameworks has been shown to be poor at times. 
\vspace{-2mm}
\section{Decentralized Unit-Commitment}\label{sec:DUC}
In this section, we present an enhanced decentralized UC problem formulation that is derived from the asynchronous ADMM framework for constrained optimization problems proposed by Wei and Ozdaglar \cite{woz}. This approach targets a multi-agent decentralized solution to the constrained optimization problem. Each agent exchanges its local estimate of the consensus variable with a neighbor after each local solve. A message exchange between two neighbors is triggered by the local clock tick associated with that edge. In doing so iteratively, all the agents converge to a solution for the global optimization problem. In our current formulation, we rely on two-way exchange of messages by a pair of neighboring regions at each global tick unlike our previous work \cite{ramanan2017asynchronous} that used a broadcast based method. In addition to the consensus quantities themselves, we also exchange their respective Lagrangian information in order to adhere to the convergence conditions set forth in \cite{woz}. 

Our decentralized formulation for the UC problem is geared towards improving solution quality in an asynchronous setting to bolster its applicability in a real world, geographically distributed computational environment. In asynchronous computational conditions, heuristics proposed in the literature with a synchronous approach in mind (i.e. fix-and-release) might be impractical.
Further, the solution mechanism illustrated in \cite{ramanan2017asynchronous} makes it evident that decentralized asynchronous methods offer good computational potential but leave a lot of room for improvement in terms of the solution quality owing to oscillations from the binary components of the UC problem. 

A region based decomposition partitions the set of buses such that each bus is uniquely owned by only one region. It follows that every bus in a region can be categorized either as a boundary or an internal bus. Boundary buses of a region are ones which have a transmission line connecting them to at least one bus belonging to a neighboring region.
Further, buses owned by a neighboring region lying on the other end of a transmission line from a boundary bus are termed as foreign buses. On the other hand, all buses owned by a region which have no transmission lines connecting to foreign buses are termed as internal buses. Transmission lines are identified using a unique identifier representing the buses at either end.


We present the decentralized UC objective function in Problem \eqref{eq:OPT}, the model and the constraints are listed in \eqref{eq:OPT_form}.

\begin{equation}\label{eq:OPT}
\begin{aligned}
& \mathcal{L}_r(\bm{\bar{\theta}},\bm{\bar{F}}, \bm{\lambda}, \bm{\phi}) = \sum\limits_{ t \in T}\sum\limits_{g \in G_r}d^gy^g_t + c^gx^g_t \\
& \qquad \qquad \qquad + \sum\limits_{ t \in T}\sum\limits_{g \in G_r} S^g_U\pi^g_{Ut} +S^g_D\pi^g_{Dt} \\
& \qquad \qquad \qquad + \sum\limits_{t \in T}\sum\limits_{b \in \mathcal{B}_{r}}	\left[\lambda^{b}_t		|\theta^b_t - \bar{\theta}^b_t|	+	\frac{\rho_{\theta}}{2} (\theta^b_t - \bar{\theta}^b_t)^2 \right] \\
& \qquad \qquad \qquad  + \sum\limits_{t \in T}\sum\limits_{u \in \mathcal{U}_{r}}\sum\limits_{v \in \mathcal{V}^u_{r}}\Big[\phi^{uv}_t|f^{uv}_t -\bar{f}^{uv}_t|\\
&\qquad \qquad \qquad + \frac{\rho_{f}}{2} (f^{uv}_t - \bar{f}^{uv}_t)^2 + \frac{\rho_{f}}{2} (f^{uv}_t - \tilde{f}^{uv}_t)^2\Big]
\end{aligned}
\end{equation}

\begin{subequations}\label{eq:OPT_form}
\allowdisplaybreaks[1]
\centering
\begin{align}
\underset{\bm{\theta},\bm{f},\bm{x},\bm{y}}{\text{min}} & \ \mathcal{L}_r(\bm{\bar{\theta}},\bm{\bar{F}}, \bm{\lambda}, \bm{\phi}) & & \\
\text{s.t.} & \ P^g_{min}x^{g}_t \leq y^g_t \leq P^g_{max}x^{g}_t,
\quad \forall t \in T, \forall g \in G_r & \label{eq:sq1}\\[1mm]
&-\pi^g_{Dt} \leq x^g_t -x^g_{t-1} \leq \pi^g_{Ut}, 
\ \forall t \in [2,T], \forall g \in G_r & \label{eq:sq2}\\[1mm]
&-R^g \leq y^g_t - y^g_{t-1} \leq R^g,  
\ \forall t \in [2,T], \forall g \in G_r & \label{eq:sq3}\\[1mm]
\begin{split} & \Gamma^{uv}(\theta^u_t - \theta^v_t) = f^{uv}_t, \ \forall u \in \mathcal{U}_{r}, \forall v \in \mathcal{V}^u_{r}, \forall t \in T \end{split} & \label{eq:sq4}\\[1mm]
\begin{split} & -F^{uv}_{max}\leq \Gamma^{uv}(\theta^u_t - \theta^v_t) \leq F^{uv}_{max}, 
\ \forall u \in \mathcal{U}_{r} \cup \mathcal{I}_r, \\
& \qquad \qquad \qquad \qquad \qquad \qquad \qquad \quad \forall v \in \mathcal{B}^u_{r},\forall t \in T 
\end{split} & \label{eq:sq5}\\[1mm]
\begin{split} & \sum\limits_{\forall g \in G^u_{r}}y^g_t - \delta^{u}_t = \sum\limits_{\forall v \in \mathcal{B}^u_{r}}[\Gamma^{uv}(\theta^{u}_t - \theta^{v}_t)], \  \forall u \in \mathcal{U}_{r} \cup \mathcal{I}_r,
\\& \qquad \qquad \qquad \qquad \qquad \qquad \qquad \qquad \quad \forall t \in T \end{split}& \label{eq:sq7}\\[1mm]
\begin{split} & \sum\limits_{\forall i\in U_t} \pi^g_{Ui} \leq x^g_t \leq 1-\sum\limits_{\forall i \in D_t} \pi^g_{Di}, \ \forall t \in T, \forall g \in G_r
\\ & \qquad \ U_t=[t-M^g_U+1,t], D_t = [t-M^g_D+1,t] \end{split} \label{eq:sq8}
\end{align}
\end{subequations}

Constraint \eqref{eq:sq1} ensures production at each generator being bounded by its minimum and maximum capacity. Constraints \eqref{eq:sq2} and \eqref{eq:sq8} enforce minimum up and down time for each generator. Constraint \eqref{eq:sq3} ensures generators adhere to their respective ramping limitations. Equation \eqref{eq:sq4} ensures that flow across inter-regional links is a function of the respective phase angles. Constraint \eqref{eq:sq5} enforces transmission line capacity constraints. Equation \eqref{eq:sq7} ensures that at each bus the 
demand is met by either power generated by attached generators, or with the flow into the region. Equations \eqref{eq:sq4}-\eqref{eq:sq7} enforce global network flow constraints.

We estimate two important quantities, \emph{intermediate flows} $\bar{F}^{uv}_t$ and \emph{intermediate phase angles} $\bar{\theta}^{b}_t$ using Equation \eqref{eq:exchg} based on values received from neighboring region $r' \in \mathcal{N}_r$. In each update, region $r'$ sends flow dual values $\tilde{\phi}^{uv,r'}_t,\ \forall u \in \mathcal{U}_r, \forall v \in \mathcal{V}^u_r\cap\mathcal{U}_{r'}, \forall t \in T$ and phase angle and phase dual estimates $\tilde{\theta}^{b,r'}_t, \tilde{\lambda}^{b,r'}_t  \forall b \in \mathcal{B}_r \cap \mathcal{B}_{r'}$. 
\begin{subequations}\label{eq:exchg}
\begin{align}
\allowdisplaybreaks[1]
\hat{\lambda}^b_t & = \frac{-1}{2}(\lambda^{b}_t  + \tilde{\lambda}^{b,r'}_t) + \frac{\rho_{\theta}}{2}(\theta^{b}_t - \tilde{\theta}^{b,r'}_t)\\
\bar{\theta}^b_t & = \frac{1}{\rho_{\theta}}(\hat{\lambda}^b_t + \lambda^{b}_t) + \tilde{\theta}^{b,r'}_t, \quad \lambda^{b}_t = \hat{\lambda}^b_t \\
\allowdisplaybreaks[3]
\hat{\phi}^{uv}_t & = \frac{-1}{2}(\phi^{uv}_t  + \tilde{\phi}^{uv,r'}_t) + \frac{\rho_f}{2}(f^{uv}_t - \tilde{f}^{uv,r'}_t)\\
\bar{f}^{uv}_t & = \frac{1}{\rho_f}(\hat{\phi}^{uv}_t + \phi^{uv}_t) + \tilde{f}^{uv,r'}_t, \quad
\phi^{uv}_t  = \hat{\phi}^{uv}_t 
\end{align}
\end{subequations}
In order to avoid redundant and expensive communication, the flow estimates $\tilde{f}^{uv}_t$ of the neighbor region are computed based on the phase angles sent by the neighbor.

\section{Asynchronous Solution Methodology}\label{sec:sm}
In this section, we seek to design a solution methodology for Problem \ref{eq:OPT_form} that performs well in asynchronous conditions and maintains operational privacy. We augment the existing formulation with a redundant valid inequality based on production and demand to boost computational performance in an asynchronous system. The two-way message exchange necessitated by local clock tick as mentioned in \cite{woz} and explained in Section \ref{sec:DUC} is tedious to implement and may cause significant computational overhead. Therefore, we introduce the concept of a controller that matches two neighboring regions on the completion of their respective local computation step. We design and develop the controller mechanism to also track asynchronous global progress of the regions while preserving asynchronous convergence conditions. 
Finally, we present our two phase asynchronous decentralized UC algorithm that solves the local convex relaxations of Problem \ref{eq:OPT_form} in the first phase before imposing binary constraints in the second phase. In order to improve computational speedup, our algorithm incorporates a novel interleaved binary mechanism that lets regions advance to their local binary problems based on consistent local convergence of their convex relaxations.
\vspace{-2mm}
\subsection{A privacy preserving valid inequality}
In a decentralized UC solution, global production has a direct bearing on the binary commitment variables. Theoretically, network flow constraints ought to be sufficient conditions for the UC problem solution to balance global production and demand. However, in a decentralized environment, network flow constraints are enforced using Lagrangian decomposition between regions. Decomposed network flow constraints drive the optimal assignment of binary decision variables which ultimately culminates in global convergence. Owing to volatility arising out of heavy latency induced message passing in an asynchronous system, the network flow constraints alone might not be strong enough to meet this balance. The decentralized formulation must therefore be further secured by a globally redundant constraint based on production and demand. These constraints must also retain the privacy preserving nature of decentralized methods. 

We consider a balance of global production and demand denoted by $\sum\limits_{\substack{\forall r \in \mathcal{R}}}\sum\limits_{\substack{\forall b \in \mathcal{U}_r\bigcup\mathcal{I}_r}}\delta^b_t = \sum\limits_{\substack{\forall r \in \mathcal{R}}}\sum\limits_{\substack{\forall g \in G_r}} y_t^g$.
In order to decentralize the production-demand balance constraint, we establish an asynchronous friendly mechanism in order to enforce it globally. We respectively compute the local production difference $\psi_{r,t}$, the inverse of the average production residual cost $s_{r,t}$ and its multiplier $\mu_{r,t}$ for every region $r$ and for every time period in the planning horizon as follows.
\begin{subequations}\label{eq:pdm}
\begin{align}
\psi_{r,t} & = \sum\limits_{\forall b \in \mathcal{U}_r\bigcup\mathcal{I}_r}\delta^b_t - \sum\limits_{\forall g \in G_r}y_t^g\\
s_{r,t} & = \frac{\sum\limits_{\forall g \in G_r} (P^g_{max} - y_t^g) }{\sum\limits_{\forall g \in G_r} d^g (P^g_{max} - y_t^g) }\\
\mu_{r,t} & = \frac{s_{r,t}}{\sum\limits_{\forall r \in \mathcal{R} }s_{r,t}}
\end{align}
\end{subequations}
Then, the local production target is then computed as follows.
\begin{equation}\label{eq:prod_diff}
\bar{p}_{r,t} =  \sum\limits_{\forall g \in G_r} y_t^g + \mu_{r,t}\sum\limits_{\forall r \in \mathcal{R}} \psi_{r,t}
\end{equation}
Therefore, Problem \eqref{eq:OPT} is further augmented by Lagrangian penalty terms pertaining to production difference as described in Problem \eqref{eq:OPT_aug}.
\begin{subequations}\label{eq:OPT_aug}
\allowdisplaybreaks[1]
\begin{align}
\ \underset{\bm{\theta},\bm{f},\bm{x},\bm{y}}{\text{min}} & \qquad \mathcal{L}_r(\bm{\bar{\theta}},\bm{\bar{F}}, \bm{\lambda}, \bm{\phi}) \\ & + \sum\limits_{t \in T}\Big[\eta_t |p_{r,t} -\bar{p}_{r,t}| + \frac{\rho_p}{2} (p_{r,t} - \bar{p}_{r,t})^2 \Big] \nonumber \\[1mm]
\text{s.t.} & \qquad \eqref{eq:sq1}-\eqref{eq:sq8} \nonumber\\[1mm]
& \qquad \sum\limits_{\forall g \in G_r}y^g_t = p_{r,t}, \quad \forall t \in T & \label{eq:sq9}
\end{align}
\end{subequations}

We further add Equation \eqref{eq:sq9} to the existing constraint set where we try to provide a production target for Problem \eqref{eq:OPT_aug}. This is achieved by performing a global weighted average of the production difference arising out of every region. Intuitively, each region is assigned a customized production target as given by $\bar{p}_{r,t}$ further strengthening convergence. It is important to note that computation of $\bar{p}_{r,t}$ is highly suited for an asynchronous model owing to a reduced volatility in values due to the multiplier $\mu_{r,t}$.
\vspace{-2.5mm}
\subsection{Compute Architecture}
One of the most important conditions imposed by the asynchronous algorithm in \cite{woz} is the presence of a global clock that drives two-way exchange of messages among agents \cite{woz}. At each global clock tick a pair of neighboring agents are triggered and exchange local information with each other leading to a two-way message exchange paradigm referred to as a doubly stochastic system. 

The requirement of double stochasticity can prove to be a limitation for a variety of reasons. From a practical standpoint, especially in a geographically distributed computational setup, the implementation of a global clock is tedious and leads to a heavier computational burden with lesser accuracy \cite{kopetz1987clock}. 
From a computational perspective, techniques relying on a doubly stochastic system also suffer from issues related to potential bottlenecks in case of compute node failure \cite{tsianos2012consensus}. If not given designed in the right manner, doubly stochastic systems can undermine the benefits of a decentralized method.

In order to solve a decentralized asynchronous constrained optimization problem, it is necessary to comply with the convergence conditions proposed in \cite{woz} and simultaneously address the issues arising out of a doubly stochastic algorithm. Therefore, in this paper we propose the concept of an additional computational agent called a controller. The controller for a doubly stochastic asynchronous decentralized scheme has the following roles:
\begin{itemize}
\item facilitating exchange of messages between neighboring regions on each clock tick.
\item helping detect global convergence of the algorithm for an asynchronous method.
\item computing an estimate of a global sum asynchronously.
\end{itemize} 

\begin{algorithm}
\caption{Controller Logic}\label{alg:ctrl}
\begin{algorithmic}
\State initialize $\bm{\tilde{\psi}}^r, \bm{\tilde{s}}^r,  \tilde{\xi}^r,\tilde{\kappa}^r \leftarrow 0, \forall r \in \mathcal{N}$
\While{GC = false}
	\State recv $\{\bm{\tilde{\psi}}^{r_1},\bm{\tilde{s}}^{r_1},\tilde{\xi}^{{r_1}},\tilde{\kappa}^{r_1}\}$ from some region $r_1$
	\If{$\tilde{\xi}^{r_2} = 1$, such that $\exists r_2 \in \mathcal{N}_{r_1}$}
		\State send to $r_1$ $\{\sum\limits_{r=1}^{|\mathcal{R}|} \bm{\tilde{\psi}}^r,\sum\limits_{r=1}^{|\mathcal{R}|}\bm{\tilde{s}}^r,\sum\limits_{r=1}^{|\mathcal{R}|} \tilde{\xi}^r,r_2\}$
		\State send to $r_2$ $\{\sum\limits_{r=1}^{|\mathcal{R}|}\bm{\tilde{\psi}}^r,\sum\limits_{r=1}^{|\mathcal{R}|} \bm{\tilde{s}}^r,\sum\limits_{r=1}^{|\mathcal{R}|} \tilde{\xi}^r,r_1\}$
	\EndIf
	\State if $\tilde{\xi}^{r},\tilde{\kappa}^{r} =1, \forall r \in \mathcal{R}$, then GC $\leftarrow$ true
\EndWhile 
\end{algorithmic}
\end{algorithm}
Algorithm \ref{alg:ctrl} introduces the logic behind the asynchronous controller. The controller maintains a running list across the planning horizon of the global production difference vector $\bm{\tilde{\psi}}^r$, the global inverse average residual cost vector $\bm{\tilde{s}}^r$, the local convergence values $\tilde{\xi}^r$ and the phase $\tilde{\kappa}^r$ of each region. As soon as a region finishes its local computation, it sends its updated local production residual as well as the inverse average residual cost,  its local convergence value and its phase to the controller. The controller updates its running estimate of these values and tries to match the aforementioned region with any of its neighbors that has also completed its local computation thereby preserving the concept of a global clock. If a match is found, the latest global running estimates are communicated to the matched pair. If no other neighboring region is active, the region simply waits until it hears back from the controller. GC denotes global convergence which occurs when all regions have converged with respect to the binary phase.
Any private infrastructural data of the regions remain opaque to the controller, since production difference and the multiplier values shield any private information local to the regions. 
\vspace{-2mm}
\subsection{Asynchronous Decentralized UC Algorithm}
The convex relaxation of the UC problem plays a key role in obtaining good solution quality \cite{javad,ramanan2017asynchronous}. It is also clear that UC problems in general have a very good convex relaxation\cite{ostrowski2012tight}.  Therefore, in this paper, we focus on improving the overall solution quality and computational performance of the asynchronous decentralized UC problem by first strengthening the performance of the convex relaxation with the help of the framework proposed in \cite{woz}.

We divide the problem into two phases pertaining to the convex phase followed by the binary phase. In each phase, the Lagrangian values are also exchanged in addition to the phase angles and flow information, as dictated by \cite{woz}, in order to obtain good solution quality. The solution from the convex phase is then used as a starting point for solving the binary phase. 
\begin{algorithm}
\caption{Interleaved Binary Asynchronous Decentralized UC (ADUC) Algorithm}\label{alg:asyncd}
\begin{algorithmic}
\For{$r = 1,2,3 \ldots |\mathcal{R}|$} 
\State Initialize $\bm{\bar{\theta}}_0,\bm{\bar{f}}_0,\bm{\bar{F}}_0, \bm{\lambda}_0, \bm{\phi}_0, \bm{x}_0, \bm{y}_0,k \leftarrow 0$
\State $\kappa \leftarrow 0$, set starting phase to convex
\While {GC = false}
	\If{$(||\bm{\theta}_k - \bm{\tilde{\theta}}_k||<\alpha) \text{ and } (|| \bm{\tilde{\theta}}_k- \bm{\tilde{\theta}}_{k-1}||<\beta)$}
        \State set $\xi_{k} \leftarrow$ 1
        \State if $\xi_i = 1,\ \forall i\in [k-\zeta,k]$, then set $\kappa \leftarrow 1$
	\EndIf
	\State $\bm{\theta}_{k+1},\bm{f}_{k+1},\bm{x}_{k+1},\bm{y}_{k+1}$ calculated by Problem \eqref{eq:OPT_aug}
	\State calculate $\bm{\psi},\bm{s},\bm{\mu}$ using Equation \eqref{eq:pdm} 
    \State send $\{\bm{\psi,s},\xi_k,\kappa\}$ to the controller
	\State recv $\{\sum\limits_{r=1}^{|\mathcal{R}|} \bm{\tilde{\psi}}^r,\sum\limits_{r=1}^{|\mathcal{R}|}\bm{\tilde{s}}^r,\sum\limits_{r=1}^{|\mathcal{R}|}{\tilde{\xi}}^{r},r'\}$ from controller
    \If{$\sum\limits_{r=1}^{|\mathcal{R}|}\tilde{\xi}^{r} = |\mathcal{R}|$}
		\State if $\kappa = 1$, set GC $\leftarrow$ true, otherwise $\kappa \leftarrow 1$
	\EndIf
    \State compute $\bm{\bar{p}}_{r} $ using Equation \eqref{eq:prod_diff}
	\State send tuple $\Delta^{ADUC}_{rr'} = \{\bm{\Theta},\bm{\Lambda}, \bm{\Phi}\}$ to $r'$ 
	\State recv tuple $\Delta^{ADUC}_{r'r} = \{\bm{\tilde{\Theta}}_{r'},\bm{\tilde{\Lambda}}_{r'}, \bm{\tilde{\Phi}}_{r'}\}$ from $r'$
	\State compute $\bm{\tilde{f}}_{r'}$ based on $\bm{\tilde{\Theta}}_{r'}$
	\State update $\bm{\bar{\theta}}_{k+1},\bm{\lambda}_{k+1},\bm{\bar{f}}_{k+1},\bm{\phi}_{k+1}$, using Equation \eqref{eq:exchg}
    \State $\eta^t_r = \eta^t_r + \rho_p(p_{r,t} - \bar{p}_t),\quad \forall t\in T$
	\State $k\leftarrow k+1$
\EndWhile
\EndFor
\end{algorithmic}
\end{algorithm}

\begin{figure}[!ht]
\begin{minipage}{0.5\textwidth}
\centering
\includegraphics[width=\textwidth,keepaspectratio]{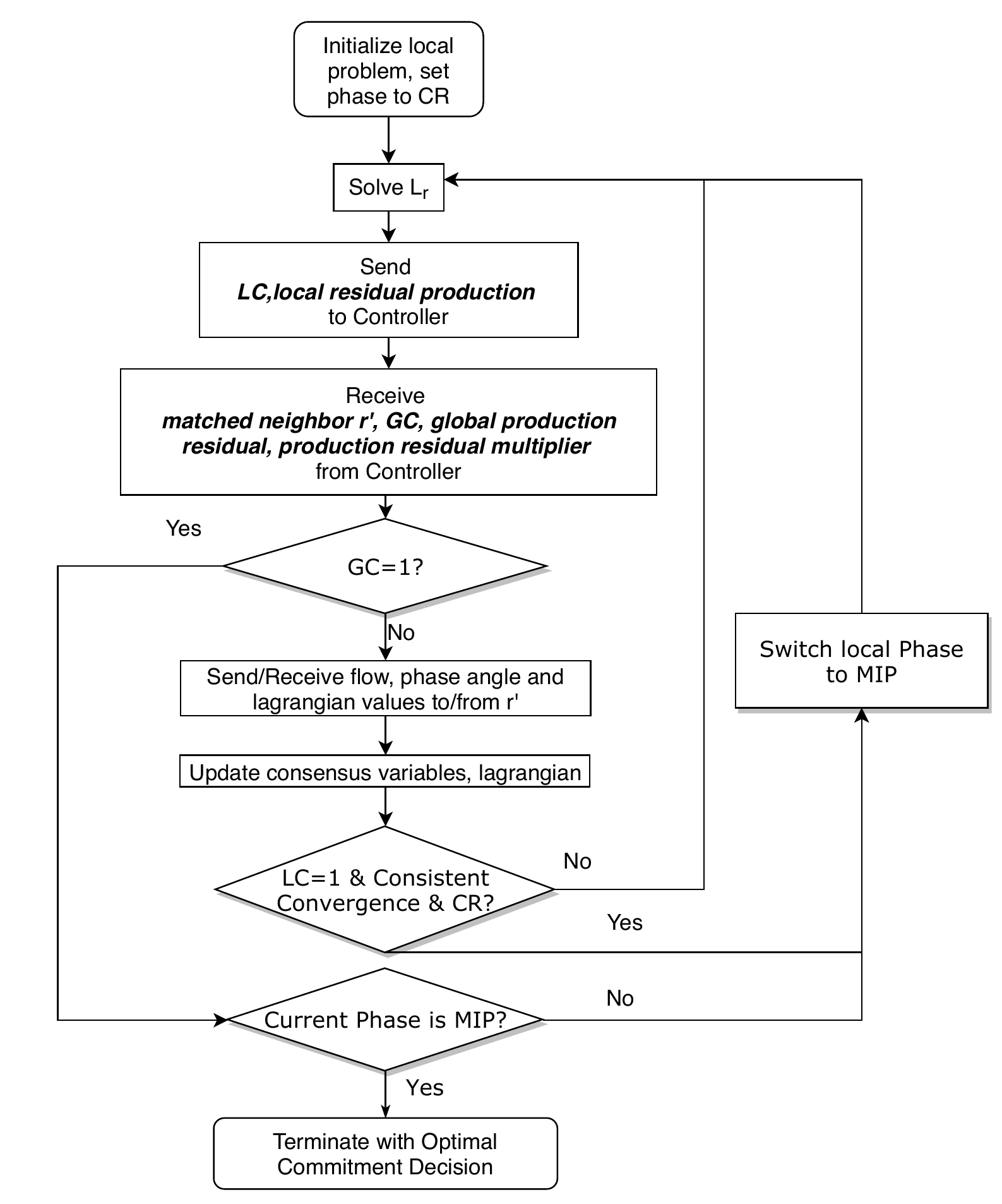}
\caption{Flowchart representing Algorithm \ref{alg:asyncd}.}\label{fig:flowchart}
\end{minipage}
\end{figure}
The Interleaved Binary Asynchronous Decentralized Unit Commitment (IBAD-UC) solution methodology is presented in Algorithm \ref{alg:asyncd}. At every iteration k, each region solves its own local subproblem and generates the commitment decisions $\bm{x}_k$, dispatch decisions $\bm{y}_k$, local phase angle values $\bm{\theta}_{k}$, flow values $\bm{f}_{k}$. The region communicates the updated local production difference values and the inverse average residual cost to the controller and waits for a response. A response from the controller provides an estimate of the global production and the multiplier vector $\sum\limits_{r=1}^{|\mathcal{R}|} \bm{\tilde{\psi}}^r,\sum\limits_{r=1}^{|\mathcal{R}|}\bm{\tilde{s}}^r$ which are used in the next iteration of the local subproblem. The controller also identifies the neighbor with which the region must perform a two-way message exchange. This step initiates an exchange of information denoted by the tuple $\Delta^{ADUC}_{rr'}$ between region $r$ and its neighbor $r'$, where,
\[ \Delta^{ADUC}_{rr'} =\left\{
					\begin{array}{lr}
							\{\bm{\Theta}_r,\bm{\Lambda}_r\} = \{\{\bm{\theta}^{b},\bm{\lambda}^{b}\} |\forall b \in \mathcal{B}_r \cap \mathcal{B}_{r'}\}\\
       							\bm{\Phi}_r = \{\bm{\phi}^{uv}| \forall u \in \mathcal{U}_r, \forall v \in \mathcal{V}^u_r\cap\mathcal{U}_{r'}\}
       					 \end{array}
			 		\right\}
  		 		     \]
This tuple consists of the newly generated primal values as well as the Lagrangian information. After observing consistent local convex convergence behavior indicated by the Interleaved Binary (IB) constant $\zeta$, the local subproblem of the region switches from the convex relaxation to its binary counterpart.

Figure \ref{fig:flowchart} illustrates the flowchart corresponding to Algorithm \ref{alg:asyncd}. $LC$ refers to local convergence value ($\xi$). The convex relaxation phase is denoted by CR ($\kappa=0$), whereas the imposition of binary constraints on commitment variables is represented by the MIP phase ($\kappa=1$).
Local convergence occurs when the primal and dual variables with respect to the phase angles are close to some predetermined limit denoted by $\alpha$ and $\beta$ respectively. Global convergence occurs when all regions have locally converged with respect to the MIP phase.

\vspace{-2mm}
\section{Experimental Results}\label{sec:results}
We perform benchmarking studies comparing the IBAD-UC algorithm with its synchronous counterpart to demonstrate its superior computational efficiency and speed. We consider the centralized solution method in which the entire large scale UC problem is solved without region based decomposition. We benchmark the IBAD-UC algorithm with the centralized method to demonstrate comparable solution quality with a significant reduction in solution times. 

In order to demonstrate robustness of our proposed solution methodology, we show that the IBAD-UC algorithm yields consistently good quality results with respect to the 75, 100 and 120 region decompositions of the IEEE 3012 bus case. The region decompositions each consist of approximately 40, 30 and 25 buses per region on an average respectively, depicting a valid real world scenario. We consider 150 generators in the 3012 bus case that have a non trivial production capacity.

\vspace{-4mm}
\subsection{Experiment Setup}
We develop a distributed, parallel software framework that uses the MPI to orchestrate Algorithm \ref{alg:asyncd} on a high performance compute cluster. Within MPI, we rely on Remote Memory Access (RMA) paradigm for asynchronous communication. RMA windows allow remote processes to read and write their latest values. Since each region occupies one process, regions communicate with their neighbors with much simpler semantics offered by RMA.

We perform our experiments on a HPC cluster comprised of Intel Xeon compute nodes with 20 cores per node with a clock rate of 2.80GHz. Each region and the controller are assigned to one core. The \textit{mpi4py} \cite{mpi4py} framework  was used to interface with MPI to conduct our HPC simulations. \textit{Gurobi 6.5} was used to solve Problem \eqref{eq:OPT_form} locally on each core with multi threading turned off on each region to prevent oversubscription of computational resources. We used IEEE 3012 bus case data from \textit{MATPOWER} \cite{matpower} for our experiments. Our experimental results pertain to the 24 hour planning horizon spanning an entire day collected from over 400 experiments conducted with the IBAD-UC algorithm. We benchmark our results against the decentralized synchronous solution methodology presented in \cite{javad} augmented with the production difference valid inequality with the multiplier being constant $\mu_{r,t} =\frac{1}{|\mathcal{R}|}$. 
\begin{figure}[!ht]
\begin{minipage}{0.5\textwidth}
\centering
\includegraphics[width=\textwidth,keepaspectratio]{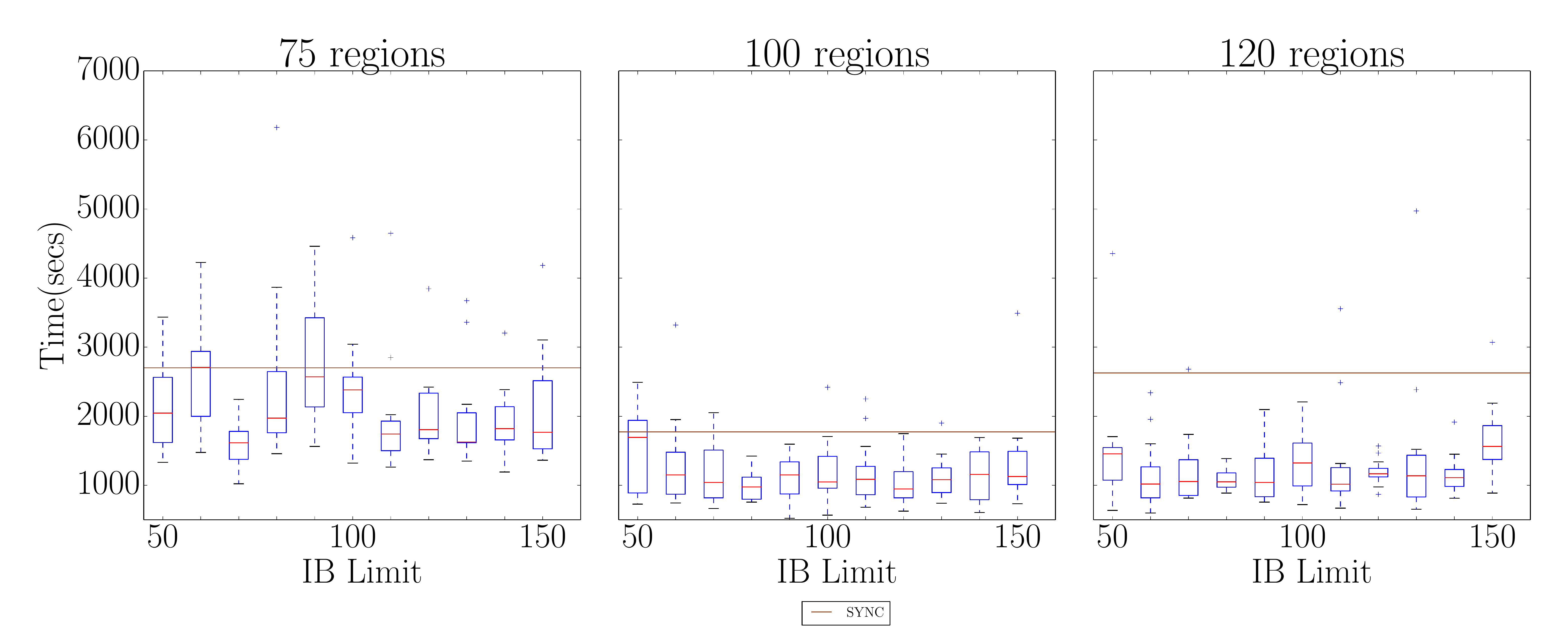}
\caption{Total time taken and the effect of IB limit $\zeta$}\label{fig:time}
\end{minipage}
\end{figure}
\begin{figure}[!ht]
\begin{minipage}{0.5\textwidth}
\centering
\includegraphics[width=\textwidth,keepaspectratio]{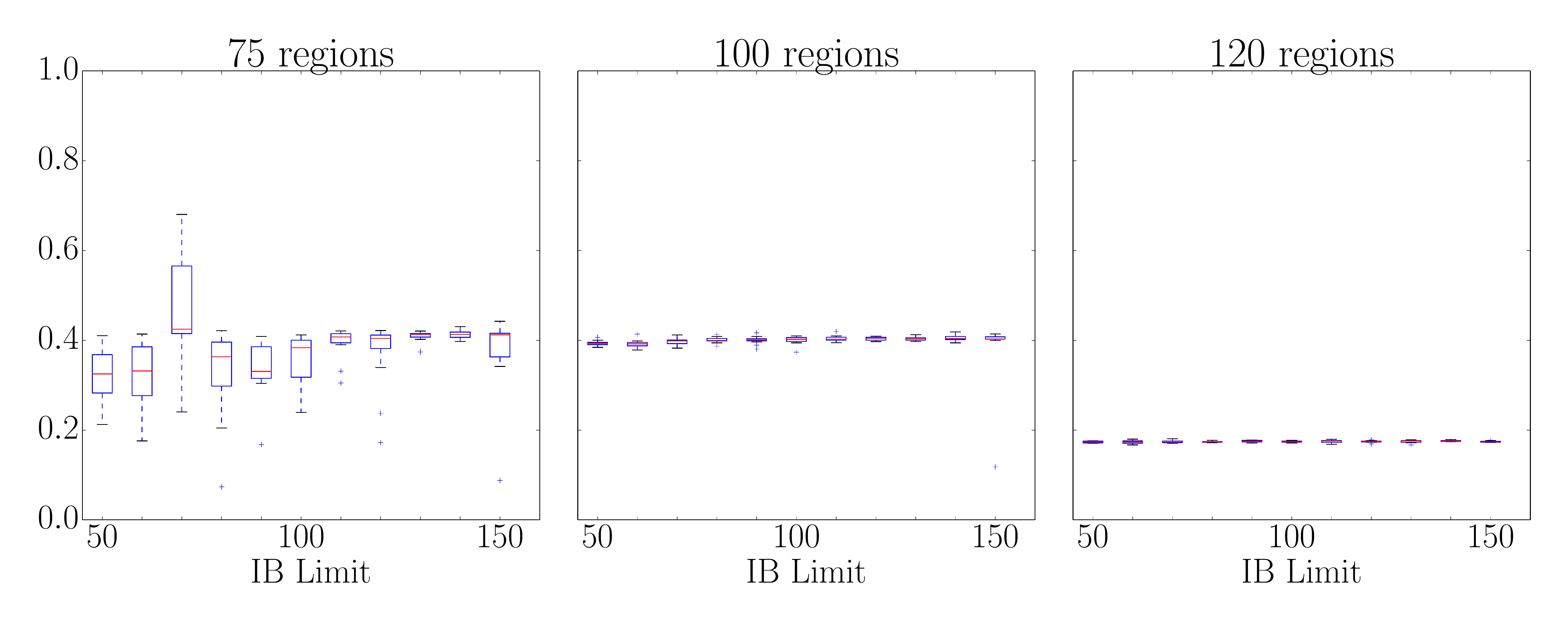}
\caption{Effect of IB limit $\zeta$ on asynchronous degree}\label{fig:ad}
\end{minipage}
\end{figure}
\begin{figure*}
\centering
\begin{minipage}{\textwidth}
        \includegraphics[width=\textwidth,keepaspectratio]{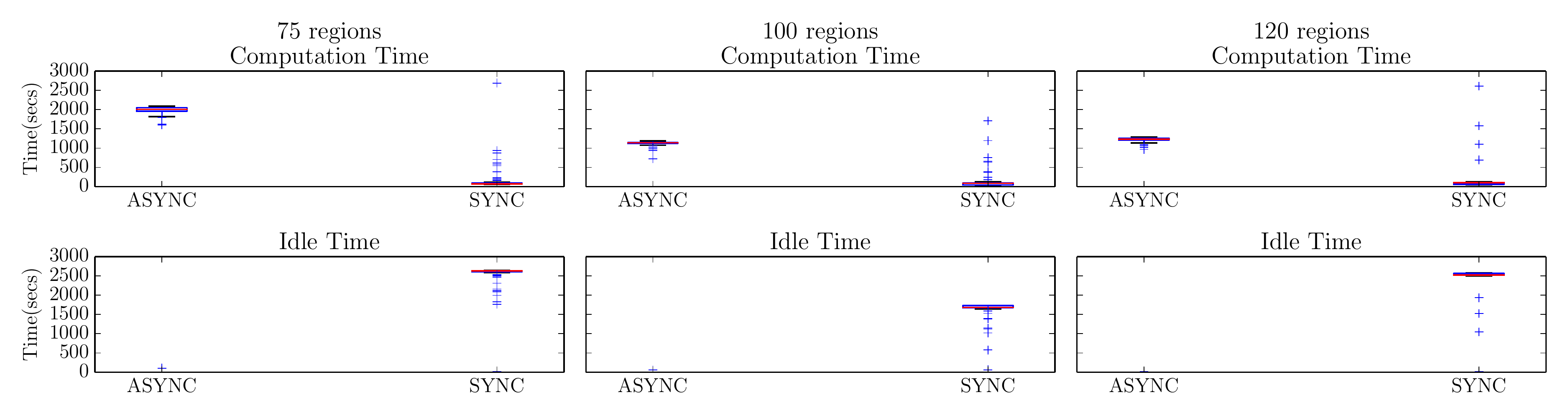}   
        \caption{Variation in Computation and Idle time for 75,100 and 120 regions}\label{fig:dcpt}
        \end{minipage}
\end{figure*}
\vspace{-2mm}
\subsection{Total solve time and effect of IB Limit}
Figure \ref{fig:time} presents box plots for the time taken for convergence by the IBAD-UC algorithm against the time taken by its synchronous counterpart. Results depict the effects of variation in the IB Limit parameter represented by $\zeta$. We observe a large variation in the time taken for convergence in the 75 region case with varying $\zeta$, whereas, for 100 and 120 region cases, the variation consistently decreases, with 120 region case being the fastest.  Figure \ref{fig:time} tells us that the 75 region decomposition likely has a relatively greater degree of imbalance in the problem sizes compared to the 100 and 120 region cases. 
It is also interesting to note that performance trends with respect to $\zeta$ oscillate between high and low variations in convergence time successively as $\zeta$ is increased, although this trend becomes much more subtle as we move from the 75 to the 120 region case.
Overall, it can be seen that despite a variation in performance with respect to $\zeta$, the IBAD-UC algorithm outperforms synchronous in all three cases.
\vspace{-4mm}
\subsection{Asynchronous Degree}
The asynchronous degree is the ratio of the minimum and the maximum number of updates performed by any region within each case. 
In Figure \ref{fig:ad} we present box plots of the asynchronous degree based on $\zeta$ as a means to measure how asynchronous the system is. We can see that the asynchronous degree shows significant variation with a lower $\zeta$ value and stabilizes as we increase $\zeta$. For the 100 and 120 region case presented in Figure  there is little variation in the asynchronous degree as $\zeta$ is changed. This observation indicates that the 75 region case is relatively imbalanced leading to variation in asynchronous degree which is exacerbated at lower $\zeta$ values indicating premature advancement into the binary phase leading to higher volatility in solution times. Figure \ref{fig:ad} thereby corroborates  Figure \ref{fig:time} since higher variation in asynchronous degree might lead to higher variations in solution times as well.
\begin{table*}
\hspace{0.01\textwidth}
	\resizebox{0.22\textwidth}{!}{\begin{minipage}[b]{0.29\textwidth}
    	\centering
		\caption{Time (secs) 75 Regions}
        \label{tab:avgtime}
    	\begin{tabular}{ |c|c|c|c|c|c|c|c|c|c|c|c|c|c|c|c| }
           \hline
           \multicolumn{1}{|>{}c|}{} & \multicolumn{4}{c|}{75 Regions} & \multicolumn{4}{c|}{100 Regions} & \multicolumn{4}{c|}{120 Regions}\\
           \cline{2-13}
           \multicolumn{1}{|>{}c|}{} & \multicolumn{2}{c|}{Asynchronous} & \multicolumn{2}{c|}{Synchronous} & \multicolumn{2}{c|}{Asynchronous} & \multicolumn{2}{c|}{Synchronous} & \multicolumn{2}{c|}{Asynchronous} & \multicolumn{2}{c|}{Synchronous}\\
           \arrayrulecolor{black}
           \cline{2-13}
           \multicolumn{1}{|>{}c|}{Time} & Mean & \% of Total & Mean & \% of Total & Mean & \% of Total & Mean & \% of Total & Mean & \% of Total & Mean & \% of Total \\
           \hline
           Comp & 1989.25 & 92.76 & 171.47 & 6.35 & 1128.81 & 94.28 & 130.61 & 7.36 & 1222.36 & 94.11 & 137.03 & 5.21\\
           Comm &1.05 & 0.05 & 0.87 & 0.03 & 0.76 & 0.06 &  0.65 & 0.04 & 1.14 & 0.09 & 0.87 & 0.03\\
           Idle  & 153.62 & 7.16 & 2528.97 & 93.61 & 67.42 & 5.63 & 1642.65 & 92.58 & 74.87 & 5.76 & 2489.31 & 94.73\\
           \hline
           Total & 2144.43 & - & 2701.43 & - & 1197.32 & - & 1774.23 & - & 1298.80 & - & 2627.67 & - \\
           \hline
        \end{tabular}
    \end{minipage}}
\end{table*}
\vspace{-4mm}
\subsection{Average Times}
Table \ref{tab:avgtime} presents the average computation, communication and idle times incurred by the asynchronous and the synchronous methods for the 75, 100 and 120 region cases. 
Figure \ref{fig:dcpt} presents the variation in the mean computation and idle times for every region incurred by IBAD-UC alongside those for the synchronous method.

From the tables, we observe that the average synchronous computation time in general is much lower than the asynchronous while also incurring a smaller percentage share of the total as well. While the asynchronous method spends relatively less time idling, the synchronous method suffers from greater amount of idle time, both in terms of average and the percentage times. In addition, the communication times and the respective percentages do not depict much variation between asynchronous and synchronous methods. The consistency observed in terms of computation, communication and idle times by various region decompositions show robustness in computational performance by the IBAD-UC algorithm.

Figure \ref{fig:dcpt} provides deeper insight into the trends presented in Table \ref{tab:avgtime} by presenting \textit{regional variations} in computation and idle times. We observe that there is wide variation in computation and idle times for regions in the synchronous method. For the synchronous method, the highest computation time incurred by a region is also very similar in value to the highest idle time incurred by any region for all cases. However, in the asynchronous method, the lower and upper bounds on computation times are much tighter and idle times are negligible. This behavior in computation and idle time is observed uniformly across all the region decompositions. 

The data presented in Figure \ref{fig:dcpt} and Table \ref{tab:avgtime} indicates that the higher computation time for a few regions form the main bottlenecks for global progress which are readily circumvented by asynchronous methods. Meanwhile, global computational progress in the synchronous method is held up by the slowest region which simultaneously incurs very high idling times on the fastest region. Despite strongly asynchronous systems, more frequent asynchronous updates are able to successfully drive the problem towards the global solution much faster leading to superior computational efficiency.
\vspace{-2mm}
\subsection{Solution Quality}
We solve the centralized problem with a 0.1\% MIPGAP, the lower bound of which is used to compute the worst case benchmarking for the IBAD-UC algorithm solution quality. We denote $\gamma$ to be the total optimal objective value comprised of operations and commitment components. $\gamma_{async},\gamma_{c}$ represent the optimal objective for the asynchronous method and the centralized method respectively.
\begin{table}
	\resizebox{0.22\textwidth}{!}{\begin{minipage}[b]{0.29\textwidth}
        \centering
        \caption{Centralized Solution}\label{tab:sq}
            \begin{tabular}{ |c|c|}
             \hline
             Total Objective ($\gamma_c$) & 108316.6\\
             \hline
             Lower Bound ($\floor{\gamma_c}$) &  108218.5\\
             \hline
             Time (secs) & 19139 \\ 
             \hline
        \end{tabular}
	\end{minipage}}
	\resizebox{0.22\textwidth}{!}{\begin{minipage}[b]{0.29\textwidth}
        \centering
        \caption{Solution Quality}\label{tab:sq2}
            \begin{tabular}{ |c|c|c| }
             \hline
             Regions & Mean Gap (\%) & Std Dev. \\ 
             \hline
             75 & 1.891 & 0.881 \\ 
             100 & 1.305 & 0.285 \\
             120 & 1.611 & 0.122\\ 
             \hline
        \end{tabular}
     \end{minipage}}
\end{table}
We present the centralized results in Table \ref{tab:sq} where $\gamma_c$ and $\floor{\gamma_c}$ represent the associated objective and lower bound. We use this to calculate a conservative solution quality in terms of the optimality gap($\frac{(\gamma_{async} - \floor{\gamma_c})*100}{\floor{\gamma_c}}$) in order to provide the worst case optimality gap for our algorithm.
We compute the mean optimality gap among multiple runs of the asynchronous method for 75, 100 and 120 region cases. 

Table \ref{tab:sq} shows that, the asynchronous solution quality is highly consistent among the regions. The asynchronous method on an average is able to consistently solve the decentralized UC problem with less than 2\% optimality gap. 
Drawing insights from Figure \ref{fig:ad} and Figure \ref{fig:time}, it can be argued that despite a highly asynchronous system, the variation in solution times as well as the solution quality are relatively small. Further, the optimal objective costs are close to that of the centralized solution indicating a robust solution with higher computational efficiency. 

\subsection{Objective Costs}
\begin{figure}[!ht]
\centering
\includegraphics[width=0.4\textwidth,keepaspectratio]{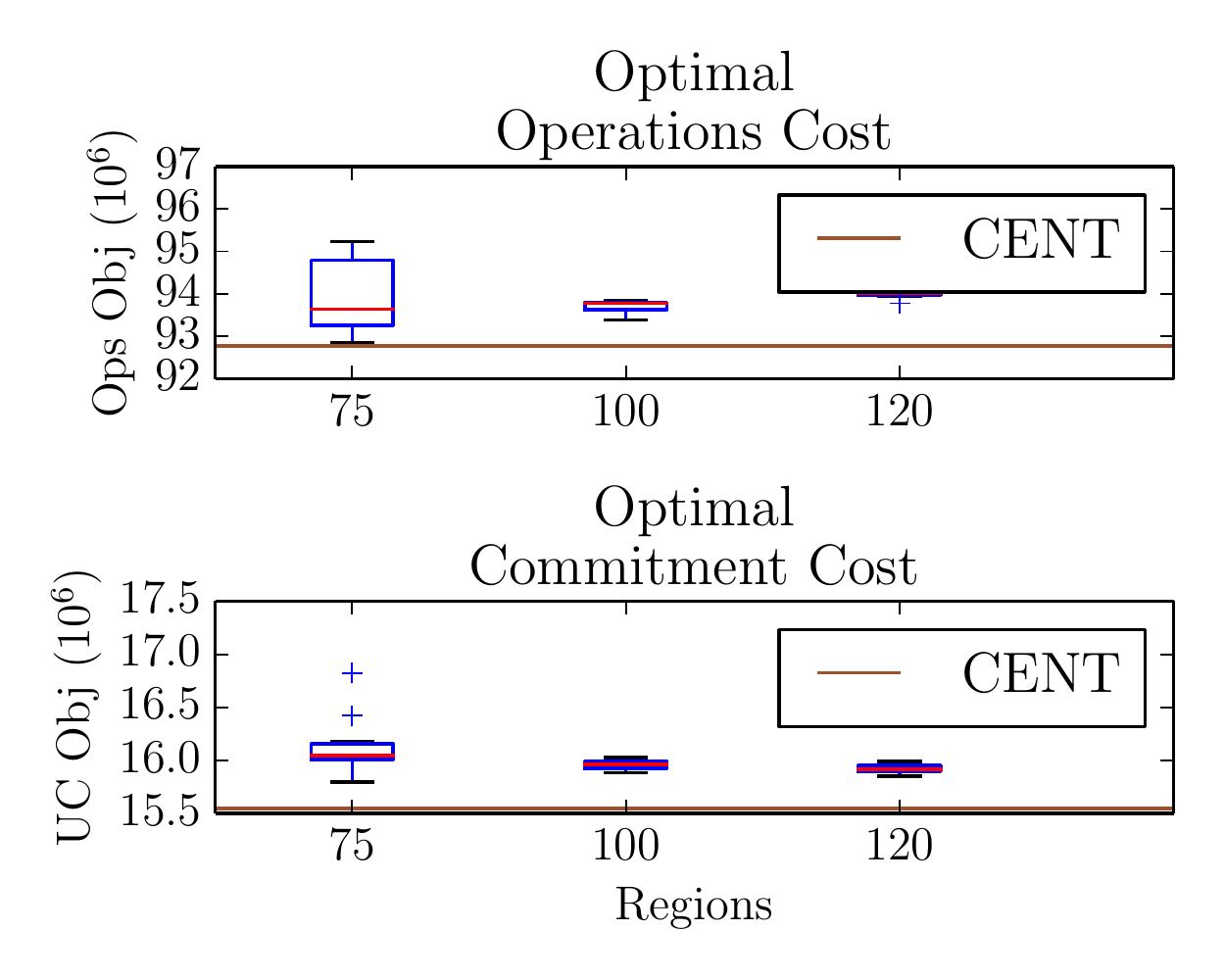}\label{fig:optcosts}
\caption{Objective costs}\label{fig:obj_costs}
\end{figure}
Figure \ref{fig:obj_costs} shows the objective cost comparison of IBAD-UC relative to the centralized solution with respect to the 75,100 and 120 region cases. We can see that the IBAD-UC yields commitment and operation decisions which are close enough to the centralized method. We can also observe the stability of the solutions within each region case owing to the valid inequalities enforced by IBAD-UC.

\vspace{-2mm}
\section{Conclusion}\label{sec:conc}
In this paper we present a novel asynchronous decentralized solution methodology for solving the UC problem for large scale power systems. Unlike other asynchronous reformulations proposed in the past that leverage a master-slave hierarchical computational model, our IBAD-UC algorithm is decentralized in nature and intended for a real-time geographically distributed heterogeneous computing environment. Our decentralized problem formulation is constructed with a strong emphasis on the privacy of region level infrastructure data and incorporates a redundant privacy preserving valid inequality. Leveraging the valid inequality the proposed asynchronous method is able to offer considerable algorithmic improvements with respect to stability and robustness of the solution. We propose a controller mechanism that implements two-way message exchanges between regions at discrete global clock ticks without a significant computational burden. 

We present HPC simulation studies based on a custom-made software framework developed by us to show the superior computational performance of the asynchronous method, along with stable solution quality. We also benchmark the asynchronous convergence characteristics with respect to the synchronous method and analyze the solution quality against that of the centralized method. Our experiments show that asynchronous methods offer a viable, robust and computationally efficient alternative to the state of the art synchronous decentralized methods. 
\vspace{-2mm}
\bibliographystyle{ieeetr} 
\bibliography{main}
\vskip 0pt plus -1fil
\begin{IEEEbiographynophoto}{Paritosh Ramanan}
Paritosh Ramanan is a 4th year PhD Candidate in Computational Science and Engineering with the School of Industrial and Systems Engineering at Georgia Institute of Technology in Atlanta, Georgia. Prior to his PhD he earned a Masters in Computer Science from Georgia State University in Atlanta, Georgia in 2015 and obtained his Bachelors in Information Systems from Birla Institute of Technology and Science (BITS) Pilani, Goa Campus in 2013. His research focuses on developing decentralized algorithms for improved computational performance of large scale optimization problems through the use of parallel and distributed computing paradigms. 
\end{IEEEbiographynophoto}
\vskip 0pt plus -1fil
\begin{IEEEbiographynophoto}{Murat Yildirim}
Dr. Murat Yildirim is an Assistant Professor in the Department of Industrial and Systems Engineering at Wayne State University. Prior to joining Wayne State, he worked as a postdoctoral fellow at the Georgia Institute of Technology (2016-2018), and obtained a Ph.D. degree in Industrial Engineering, and B.Sc. degrees in Electrical and Industrial Engineering from the same institution. Dr. Yildirim's research interest lies in advancing the integration of mathematical programming and data analytics in large scale energy systems. Specifically, he focuses on the modeling and the computational challenges arising from the integration of real-time sensor inferences into large-scale mixed integer programs (MIPs) used for optimizing and controlling networked systems.
\end{IEEEbiographynophoto}
\vskip 0pt plus -1fil
\begin{IEEEbiographynophoto}{Edmond Chow}
Edmond Chow is an Associate Professor in the School of Computational Science and Engineering at Georgia Institute of Technology.  He previously held positions at D. E. Shaw Research and Lawrence Livermore National Laboratory.  His research is in developing numerical methods specialized for high-performance computers, including asynchronous iterative methods, and applying these methods to solve large-scale scientific computing problems.  Dr. Chow was awarded the 2009 ACM Gordon Bell prize and the 2002 U.S. Presidential Early Career Award for Scientists and Engineers (PECASE).  He serves as Associate Editor for ACM Transactions on Mathematical Software and previously served as Associate Editor for SIAM Journal on Scientific Computing.
\end{IEEEbiographynophoto}
\vskip 0pt plus -1fil
\begin{IEEEbiographynophoto}{Nagi Gebraeel}
Dr. Nagi Gebraeel is the Georgia Power Early Career Professor and Professor in the Stewart School of Industrial and Systems Engineering at Georgia Tech. His research interests lie at the intersection of industrial predictive analytics and decision optimization models for large scale power generation applications. Dr. Gebraeel serves as an associate director at Georgia Tech's Strategic Energy Institute and the director of the Analytics and Prognostics Systems laboratory at Georgia Tech's Manufacturing Institute. Dr. Gebraeel was the former president of the Institute of Industrial Engineers (IIE) Quality and Reliability Engineering Division, and is currently a member of the Institute for Operations Research and the Management Sciences (INFORMS).
\end{IEEEbiographynophoto}

\end{document}